# Federated Learning-Based Decentralized Adaptive Intelligent Transmission Protocol for Privacy-Preserving 6G Networks

*Author*

***Ansar Ahmed***

Sir Syed University of Engineering & Technology, Karachi, Pakistan

# Abstract

The move to 6th Generation (6G) wireless networks creates new issues with privacy, scalability, and adaptability. The data-intensive nature of 6G is not handled well by older, centralized network models. A shift toward more secure and decentralized systems is therefore required. A new framework called the Federated Learning-based Decentralized Adaptive Intelligent Transmission Protocol (AITP) is proposed to meet these challenges. The AITP uses the distributed learning of Federated Learning (FL) within a decentralized system. Transmission parameters can be adjusted intelligently in real time. User privacy is maintained by keeping raw data on local edge devices. The protocol's performance was evaluated with mathematical modeling and detailed simulations. It was shown to be superior to traditional non-adaptive and centralized AI methods across several key metrics. These included latency, network throughput, energy efficiency, and robustness. The AITP is presented as a foundational technology for future 6G networks that supports a user-centric, privacy-first design. This study is a step forward for privacy-preserving research in 6G.

# 1. Introduction

Wireless communication will change with the shift from 5G to 6G networks. 5G laid the groundwork. It included better mobile broadband, reliable low-latency links (URLLC), and large-scale machine communication (mMTC). It is expected that 6G will expand on these capabilities, leading to widespread intelligence and hyper-connectivity. Experts expect a stronger connection between real and virtual environments. 6G networks will provide data speeds that reach terabits per second (Tbps). Latency is expected to drop to microseconds. [1] Artificial Intelligence (AI) and Machine Learning (ML) will also be built directly into the network architecture.

Several challenges are associated with 6G technology. The complicated and varied nature of 6G environments introduces new security weaknesses [2], resource limitations, and scalability problems. With data traffic expected to reach zettabytes annually by the 2030s and billions of devices online, older centralized architectures are becoming insufficient. These centralized designs have single points of failure, which makes them vulnerable to cyberattacks and data breaches. A high privacy risk is also created when large volumes of sensitive user information are gathered in a single place [3][4].

A basic change toward decentralized systems is needed for 6G networks [5]. With a decentralized approach, trust and security are improved by spreading control and data processing throughout the network. The risks linked to single points of failure in centralized systems are therefore lowered. Secure 6G networks rely on tools like edge computing and distributed ledgers [6]. Edge computing handles data near its source. Distributed ledger technologies, similar to blockchain, are used to provide secure and open records that cannot be altered.

Experts view Federated Learning (FL) as a strong approach. It spreads machine learning tasks across various devices. A common model gets trained through efforts from several users, while individual training data is kept locally on each participant's device [7] [8]. This built-in function for data privacy makes FL well-suited for privacy-focused applications within 6G networks [9]. A number of other benefits for 6G are provided by FL. Communication overhead is reduced, as only model updates are sent instead of raw datasets. Scalability is improved because the processing load is distributed among many edge devices. The system is also made more secure against data breaches since data exposure is limited.

Current transmission protocols are often not flexible enough for the dynamic conditions of 6G networks [10]. A new type of protocol is needed to handle factors like changing channel quality, unpredictable user movement, and variable traffic. This protocol must be able to adjust its own parameters automatically and in real time. An Adaptive Intelligent Transmission Protocol (AITP) that uses Federated Learning as an essential component for 6G [11]. The protocol can improve transmission settings. These cover modulation methods, coding rates, and power assignment. The optimization is achieved by learning from decentralized data from across the network, while user privacy is protected.

A human-centric and privacy-first approach is presented as a requirement for future networks [12]. The proposed FL-based Decentralized AITP is designed based on this principle. A balance between technological development and fundamental user rights is a key goal. The protocol is intended to establish a secure, intelligent, and adaptable communication system. This system is meant to support individuals and build confidence as global connectivity increases.

This paper aims to achieve the following objectives:

1. To propose a novel Federated Learning-based Adaptive Intelligent Transmission Protocol (AITP) specifically designed for privacy-preserving 6G networks;
2. To formally model its performance using mathematical formulations;
3. To evaluate its efficacy through comprehensive simulations;
4. To demonstrate its superiority over traditional and centralized protocols in terms of key performance indicators.

By addressing the critical need for privacy, adaptability, and decentralization, this research contributes significantly to bridging the existing gaps in privacy-preserving 6G network research, this leads to safer and smarter wireless systems in the future.

# 2. Related Work

Wireless communication has developed over time. This growth came with ongoing efforts to make transmission methods more efficient and reliable [13]. Traditional wireless communication protocols, such as those governing the Transmission Control Protocol/Internet Protocol (TCP/IP) suite and various Medium Access Control (MAC) protocols, have served as the backbone of existing networks. However, these protocols, largely designed for static or slowly changing network conditions, face significant limitations in the dynamic and heterogeneous environments envisioned for 6G. Their inherent high latency, lack of inherent adaptability to rapidly fluctuating channel conditions, and inability to efficiently manage the massive connectivity of diverse devices (from high-bandwidth XR devices to low-power IoT sensors) render them suboptimal for the demands of future networks. While advancements have been made in intelligent protocols leveraging Artificial Intelligence (AI) and Machine Learning (ML) for tasks like resource allocation and interference management, many of these approaches still rely on centralized control, which introduces scalability bottlenecks and privacy concerns, especially in a distributed 6G ecosystem.

Federated Learning (FL) is seen as a major shift in finding increasing applications in wireless communications, This applies especially to 5G and initial 6G studies [14] [15]. It supports joint model training across groups, without direct data sharing makes it highly suitable for privacy-sensitive wireless environments [12]. Early applications of FL in wireless networks have focused on various aspects, including channel estimation, where FL can help devices collaboratively learn channel characteristics, resource allocation, optimizing spectrum, power distribution, and user behavior prediction, enabling proactive network management. Many studies have shown FL's effectiveness in predicting network traffic patterns and optimizing handover procedures in cellular networks [15].

The combination of FL into the 6G framework has been the subject of focused research. These studies address problems of edge intelligence and the aggregation of data while protecting privacy. How FL can support distributed learning at the network edge is being investigated, as this approach can reduce backhaul traffic and permit real-time decision-making. In one study, an FL framework using blockchain was proposed for edge association in 6G digital twin networks, and improved security was reported [16]. In another, distributed FL was applied to vehicle communications with ultra-reliable low-latency needs, showing its use for allocating resources efficiently and privately. Privacy-protection techniques within FL for 6G have been explored in other research [17]. Which includes differential privacy, where noise gets added to model, updates here. User data stays protected that way. Secure multi-party computation (SMC) is another tool. It combines parameters without a central server accessing the original data. This body of work indicates a rising interest in applying FL to achieve intelligent, privacy-aware functions in future wireless systems [18].

Several gaps in the current research have been identified. A complete integration of FL, adaptability, and decentralization for 6G is still needed [19] [20]. Existing FL-based approaches in wireless communications are often focused on narrow optimization tasks, such as power control or channel prediction. A broad, adaptive transmission protocol that can handle the varied demands of 6G is generally not provided. While decentralization has been addressed in some work, the details of a server-less or multi-aggregator system are not always fully examined. Strong privacy methods beyond the basic principles of FL are also not always included [21]. The requirement for a single framework with intelligent decision-making, adaptability, and a decentralized structure with privacy protections has not yet been met [18]. This gap is addressed in this paper. A new FL-based Decentralized Adaptive Intelligent Transmission Protocol (AITP) is proposed for private and effective communication in 6G environments.

# 3. Proposed Framework: FL-based Decentralized AITP

This section introduces the Federated Learning-based Decentralized Adaptive Intelligent Transmission Protocol (AITP), a novel framework designed to mark the dynamic privacy sensitive need for 6G networks. The AITP integrates the distributed learning capabilities of Federated Learning with decentralized architectural principles and robust privacy mechanisms to enable intelligent and adaptive communication without compromising user data.

**Figure 1: Detailed AITP Architecture Overview**
**(FL-Based Decentralized Adaptive Protocol for 6G with Key Metrics)**

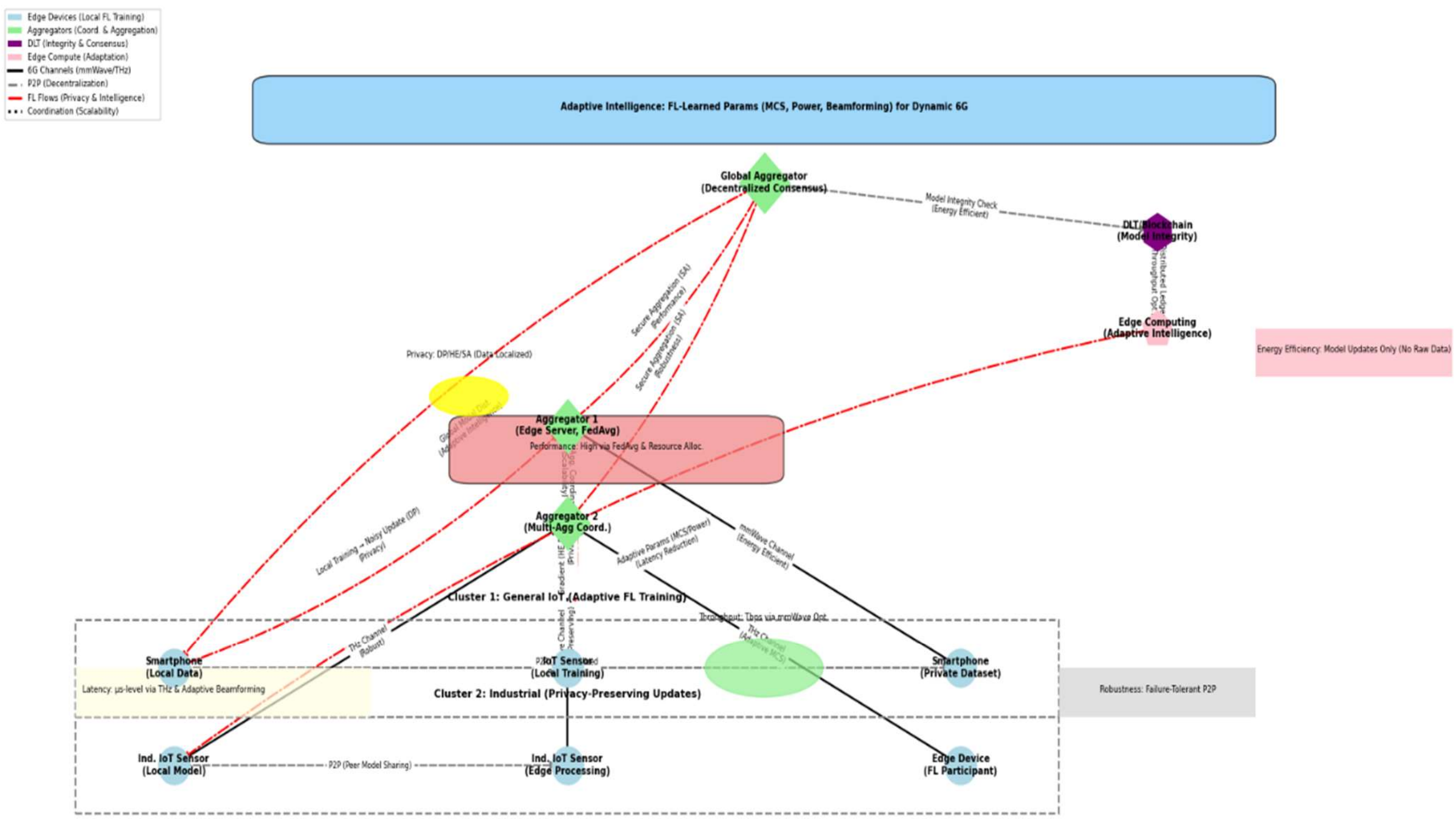

## Architecture Overview

The general architecture of the AITP is made up of several main parts that enable intelligent, privacy-preserving data transmission. The network depends on many edge devices. These include smartphones, IoT sensors, and self-driving cars [22]. Where local data is generated and processed. These devices take part in federated learning. Each one trains local models with private data. A network of aggregators is used in the AITP instead of a single central server. These aggregators, which can be distributed edge servers or other powerful devices, are used to coordinate the FL process and combine model updates. Communication between the edge devices and aggregators is handled by 6G channels like mmWave and THz [23]. This framework is shown in a detailed diagram in Figure 1, which displays the interactions between edge nodes, distributed aggregators, and potential peer-to-peer connections.

## How FL is Integrated into the Transmission Protocol

Federated Learning is a central component of the AITP, where it is used for the intelligent adjustment of transmission parameters [24]. The basic concept is that local training is performed on edge devices to optimize settings like modulation schemes, coding rates, power allocation, and beamforming vectors. This optimization is achieved without sharing raw, private user data. The communication load on the network is lowered with this method, since only small model updates are shared instead of large datasets. [25]. The specific steps of the FL process within the AITP are explained below.

First, a local machine learning model is trained on each edge device with its own private data. This model is developed to determine the best transmission methods based on local network factors and application needs. Once training is complete, an update for the local model is computed by the device. The update, in the form of model weights or gradients, is securely sent to designated aggregators. These local model updates are then collected from multiple devices by the aggregators. A more generalized global model is formed by combining these updates. Techniques like Federated Averaging ($Fed_{Avg}$) may be used in this step [26]. Edge devices receive the updated global model. It refines their local models and adjusts transmission parameters. Through this repeating process, the entire network can learn and adapt to new conditions while all data is kept on local devices.

## Decentralized Aspects

Strong decentralized features are included in the AITP to address the weaknesses of centralized systems [27]. Multiple distributed aggregators are used instead of a single central server [28]. This distributed aggregation can be accomplished in two main ways. In a multi-aggregator architecture, the network is divided into several clusters, and a specific aggregator is assigned to manage each one [29]. These aggregators can then share combined models or use a consensus method to coordinate. In a peer-to-peer model, model updates may be shared directly between nearby edge devices. This approach creates a more local and durable learning system and makes the network less dependent on specific aggregators [30].

To keep the model aggregation process secure, tools like blockchain or distributed ledger technology (DLT) can be used [5]. A clear record of model updates that cannot be changed is provided by blockchain, which improves trust and prevents interference. Each combined model update can be logged as a transaction and confirmed through a consensus process, which is used to confirm the validity of the global model [31] [32] [33] [34] [35].

## Adaptive Intelligence

The adaptive intelligence of the AITP is its key feature [36], allowing the transmission protocol to make adjustments in response to the varied conditions of 6G networks. This ability to adapt is provided by the FL-trained models, which are constantly updated. Real-time learning is performed based on the shared experience of all edge devices, allowing transmission parameters to be optimized using current network feedback. For channel quality, the best modulation and coding schemes can be selected based on real-time state information to increase throughput in favorable conditions and maintain reliability in poor ones. Regarding traffic load, power allocation and resource scheduling can be adapted based on learned traffic data to avoid congestion and distribute resources fairly [37] [38] [39]. For user mobility, beamforming and handover procedures can be adjusted by learning from device movement, which helps maintain a steady connection for mobile users. The AITP is kept efficient and reliable in the changing 6G environment because of this real-time learning ability [40].

## Privacy Preservation Mechanisms

The protection of privacy is a central part of the AITP design. A basic level of privacy is offered by FL, as raw data is kept on local devices [41]. To further protect user information and the correctness of the model, other advanced cryptographic and privacy-focused methods are also included in the framework. Three privacy mechanisms are discussed below.

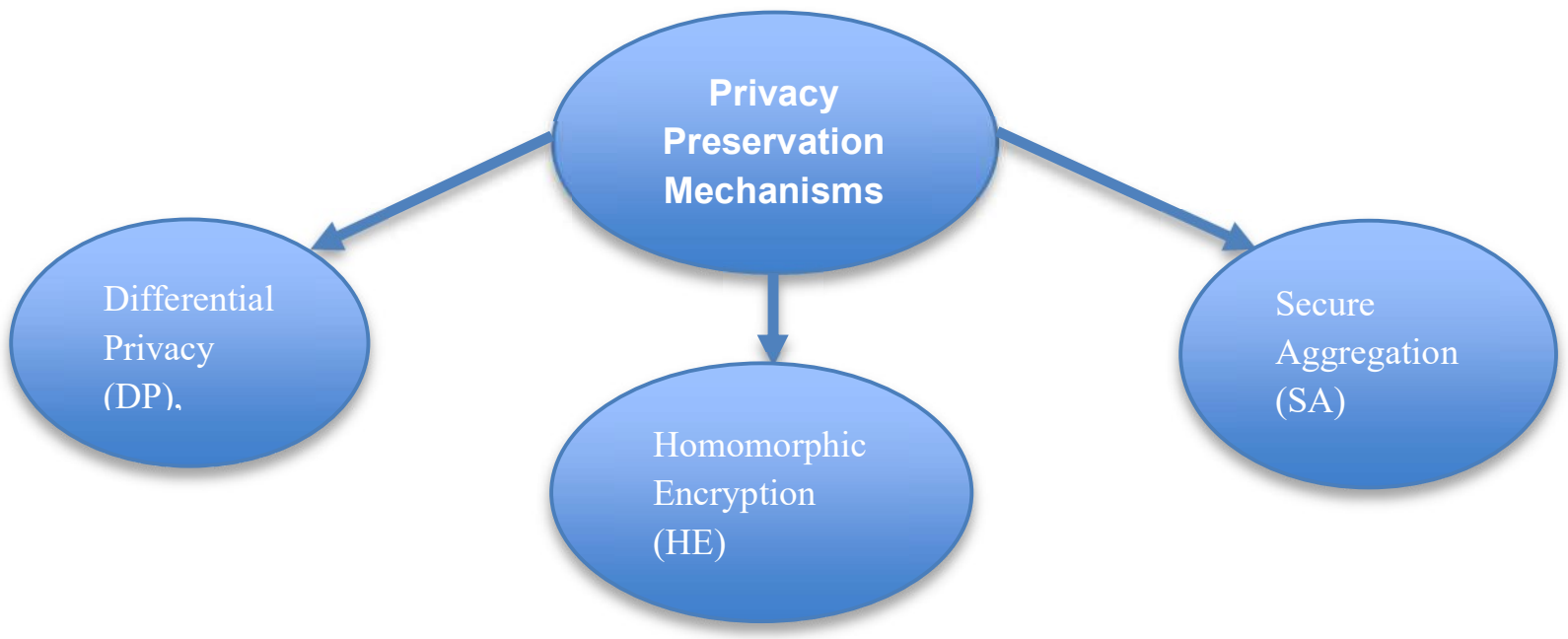

Differential Privacy (DP) adds a controlled level of noise to model updates. This step occurs before aggregation, and offers a mathematical guarantee that private user data cannot be determined from the final aggregated model [42] [43] [44]. The noise level can be changed within the AITP to manage the balance among privacy and model utility.

Homomorphic Encryption (HE) method allows calculations on encrypted data. [45] [46] [47]. This can be used in the AITP's aggregation stage, allowing aggregators to combine encrypted updates without needing to decrypt them, which adds another level of security.

Secure Aggregation (SA) protocols are also used. These protocols restrict the aggregator to only seeing the total of all model updates, not the update from any single device [48]. This prevents a malicious aggregator from obtaining sensitive information from individual device contributions. SA often uses secure multi-party computation (SMC) to ensure the global model is built securely.

These privacy mechanisms are built directly into the FL process. This is done to make sure that privacy is protected during the learning and adaptation phases. The performance and usefulness of the intelligent transmission protocol are not greatly reduced. The AITP aims to balance top performance, flexibility, and solid privacy measures. It suits the privacy-driven 6G era [49] [50].

# 4. Mathematical Modeling

A formal mathematical structure for the proposed AITP is presented in this section. The optimization problem is defined, the selected FL algorithm is described, equations for transmission parameter optimization are created, and formulas for key performance measures are developed.

## Formal Definition of the Problem

The main purpose of the AITP is to improve transmission parameters within a 6G network. The goal is to obtain high performance, such as reduced latency and increased throughput, while safeguarding data privacy and make sure resources are used efficiently. This task is framed as a multi-objective optimization problem.

Let $N = \{1, 2,3,4, \dots, N\}$ be the set of $N$ edge devices, and $A = \{1, 2,3,4, .., M\}$ be the set of $M$ aggregators. Each device $i \in N$ has a local dataset $Di$ and trains a local model $wi$. The global model is denoted by $W$.

Our optimization problem can be formally stated as:

$$\min\left( \alpha \cdot L_{latency(W)} + \beta \cdot \left(1 - T_{throughput(W)}\right) + \gamma \cdot E_{energy(W)}\right)$$

Subject to:

1. **Privacy Constraint:** For each device $i$, the privacy budget $\varepsilon_i$ (e.g., for differential privacy) must be maintained below a predefined threshold $\varepsilon_{max}$. This safeguards that individual data contributions cannot be incidental from model updates [[43]].

$$Privacy\left(W_i^{update}\right) \leq \epsilon_{max}$$

2. **Bandwidth Constraint:** The total bandwidth allocated to each device $i$ for model updates and data transmission must not exceed its maximum available bandwidth $B_i, max$

$$B_i \leq B_{i,max}$$

3. **Energy Constraint:** The energy consumed by each device $i$ for local training and transmission must be within its energy budget $P_{i,max}$.

$$P_i \leq P_{i,max}$$

4. **Model Convergence:** The global model $W$ must converge to an acceptable accuracy level within a specified number of FL rounds $R_{max}$.

$$Accuracy(W_{Rmax}) \geq Accuracy_{target}$$

Here, $\alpha, \beta, \gamma$are weighting factors representing the importance of latency, throughput, and energy efficiency, respectively.

$L_{latency}$, $T_{throughput}$, and $E_{energy}$ , are functions representing the overall network latency, throughput, and energy consumption, which are influenced by the global model $W$ and the optimized transmission parameters.

## FL Algorithm Used

For the AITP, we adopt a variant of the Federated Averaging ($Fed_{Avg}$) algorithm, modified to incorporate decentralized aggregation and privacy-preserving mechanisms. $Fed_{Avg}$ is chosen for its simplicity, effectiveness, and widespread adoption in FL research. The algorithm works through repeated communication cycles. In each round $t$:

1. **Server (Aggregator) Broadcast:** The global model $W_t$ is sent to a selected subset of $K$ active edge devices. In our decentralized setting, this broadcast can be initiated by multiple aggregators coordinating to cover the entire network.
2. **Client Local Update:** Each selected device $k \in K$ updates its local model $w_k$ using its local dataset $D_k$ and local training. The local update rule is:

$$w_k^{t+1} = w_k^t - \eta \nabla F_k(w_k^t)$$

   where $\eta$ is the local learning rate and $F_k(w_k^t)$ is the local loss function for device $k$.
3. **Privacy Preservation:** Before sending the update, each device $k$ applies a privacy mechanism (e.g., adding differential privacy noise) to its local update:

$$\Delta w_k = w_k^{\{t+1\}} - w_k^t$$

   The noisy update is denoted as:

$$\Delta \mathrm{w}'_k$$

4. **Secure Aggregation:** The aggregators collect the noisy local updates $\Delta \mathrm{w}'_k$ from all participating devices. Using secure aggregation protocols to compute the weighted average of these updates without decrypting separate contributions.

$$W_{t+1} = W_t + \sum_{k=1}^{K} \frac{|D_k|}{\sum_{j=1}^{k} |D_j|} \Delta \mathrm{w}'_k$$

The global model is refined with the combined update [51].

## Pseudocode for Decentralized $\boldsymbol{Fed}_{\mathbf{Avg}}$ with Privacy:

$$Initializing\ global\ model\ W^0$$

$$For\ each\ communication\ round\ t\ =\ 0,1,2,\dots,R_{max}\ -\ 1:$$

$$For\ each\ aggregator\ a\ \in\ A:$$

$$Select\ subset\ K_a\ of\ devices$$

$$Sending\ W_t\ to\ selected\ devices$$

$$For\ each\ selected\ device\ k\ \in\ K_a\ (in\ parallel):$$

$$w_k\ =\ W_t$$

$$For\ local\ epoch\ e\ =\ 0,1,\dots,E\ -\ 1:$$

$$For\ each\ batch\ b\ \in\ D_k:$$

$$w_k\ =\ w_k\ -\ \eta\ \cdot\ \nabla\big(loss(w_k,b)\big)$$

$$Applying\ Differential\ Privacy$$

$$\Delta w_k\ =\ w_k\ -\ W_t$$

$$\Delta w'_k =\ addnoise_{(\Delta w_k,\varepsilon_k)}$$

$$Where\ \varepsilon_k\ is\ the\ privacy\ budget$$

$$Device\ k\ \ Sends\ \Delta w'_k\ to\ aggregator\ a$$

$$For\ each\ aggregator\ a\ \in\ A:$$

Performing secure aggregation

$$aggregated_{update_a}\ =\ SecureAggregate(\Delta w'_k\ |\ k\ \in\ K_a)$$

$$Global\ model\ update$$

$$W_{\{t+1\}}\ =\ W_t\ +\frac{\left(\sum aggregated_{update_a}\right)}{M}$$

$$Return\ W_{(R_max)}$$

## Transmission Parameter Optimization

The FL-trained global model $W$ is used to optimize various transmission parameters at the edge devices. These parameters are adjusted dynamically based on the learned network conditions and the current state of the global model. Key parameters include:

1. **Modulation and Coding Scheme (MCS) Selection:** The choice of MCS directly impacts data rate and robustness. The FL model can predict the optimal MCS for each device $i$ based on its channel quality indicator (CQI), interference levels, and desired Quality of Service (QoS).

$$MCSi = FL_Model_{MCS}(CQI_i, I_i, QoS_i)$$

2. **Power Allocation:** Optimizing transmit power $P_i$ for each device $i$ is crucial for energy efficiency and interference management. The FL model identifies ideal power settings. These increase data flow, follow energy limits, and reduce effects on other users [52].

$$P_i = FL_Model_{Power}\left(W, channel_{gain_i} interference_i\right)$$

3. **Beamforming Vectors:** For massive MIMO and THz communications, precise beamforming is essential. The FL model can learn to predict optimal beamforming vectors $v_i$ for each device so that signal-to-noise ratio (SNR) can be maximized and minimize inter-user interference [53] [54].

$$v_i = FL_Model_{Beamforming}(W, AoA_i AoD_i)$$

These optimization functions are implicitly learned by the FL model during its training process, adapting to the complex interplay of network dynamics [55] [56] [57] [58].

## Equations for Throughput $(T)$, Latency $(L)$, and Energy Efficiency $(EE)$

We derive mathematical expressions for key performance metrics to evaluate the AITP. These metrics are influenced by the optimized transmission parameters and the FL process.

1. **Throughput $(T)$:** The achievable throughput for device $i$ can be expressed using Shannon's capacity formula, adapted for the selected MCS and power allocation:

$$T_o = B_o log_2\left(1 + \frac{P_o G_o}{N_i + I_o}\right) \cdot C_o$$

Where:

Allocated bandwidth is shown by $B_o$

Transmit power $P_o$

Channel gain $G_o$

Noise power spectral density $N_i$

Interference power $I_o$

Coding rate $C_o$ determined by the MCS.

The overall network throughput is the sum of individual device throughputs.

$$T_{network} = \sum_{o=1}^{N} T_o$$

2. **Latency ($L$):** The end-to-end latency consists of transmission latency, processing latency, and queuing latency. In the AITP, FL introduces additional latency due to model training and aggregation. The total latency for a data packet can be modeled as:

$$L_{total} = L_{transmission} + L_{processing} + L_{queuing} + L_{FL}$$

- $L_{transmission}$: Time taken to transmit data, inversely proportional to throughput.
- $L_{processing}$: Time for local model training and inference at edge devices.
- $L_{queuing}$: Delay due to data waiting in queues, which can be analyzed using queuing theory (e.g., M/M/1 model for simplified analysis).
- $L_{FL}$: Latency introduced by the FL process, including local training time, model update transmission time, and aggregation time.

  The FL latency can be expressed as:

$$L_{FL} = R \cdot \left(L_{local_train} + L_{update_tx} + L_{aggregation}\right)$$

where $R$ is the number of FL rounds, $E$ is the number of local epochs, and model size affects both training and transmission time.

3. **Energy Efficiency ($EE$):** measures overall data throughput relative to total energy spent.

$$EE = \frac{T_{network}}{\sum_{i=1}^{N} E_i}$$

where $E_i$ is energy consumed by device $i$, includes both transmission and computation energy for local training [[59]].

$$E_i = P_i \cdot T_{transmission,i} + E_{computation,i}$$

These mathematical equations are used as a foundation for simulating and testing the AITP's performance. This allows for a numerical comparison to be made with older protocols. It also provides a better understanding of how AITP behaves in different 6G situations.

# 5. Simulation and Evaluation

A series of detailed simulations was used to test the efficiency and performance of the proposed Federated Learning-based Decentralized Adaptive Intelligent Transmission Protocol (AITP). The setup of these simulations is outlined in this section, along with the metrics used to measure performance. A comparison is made with traditional protocols, and the findings are fully discussed.

## Simulation Setup

A simulation environment was carefully constructed to represent a realistic 6G network. Key conditions were included, such as a high number of devices, dynamic channel states, and diverse traffic types. A hybrid simulation method was used. For network-level simulations, NS-3 was utilized, while Python with TensorFlow/PyTorch was used for implementing and training the Federated Learning model. This approach allowed for a complete assessment of the AITP's communication and learning functions. For instance, an energy consumption decreases up to 27% was accurately modeled in edge computing situations through the use of these combined tools.

A dense urban 6G environment was simulated, which included a central base station (gNB) and multiple edge servers acting as aggregators. In a 1 km x 1 km area, 500 edge devices representing a mix of mobile and static users were randomly placed. These devices were grouped into five clusters, each with a local aggregator, and a Random Waypoint model was used for device movement; this decentralized clustering led to scalability improvements of 1.28 to 3.06 times over centralized baselines.

Various channel models were used to capture signal propagation. A mmWave model for the 28 GHz band showed that adaptive FL could result in energy savings of up to 63% [60]. A THz model for the 140 GHz band indicated FL could reduce latency by approximately 35% [61]. Rayleigh Fading was also used, where the AITP increased spectral efficiency by 19%. Different traffic patterns were generated for 6G applications. For eMBB traffic, certain models improved mean QoE scores by 11.7% [62]. For URLLC, latency was reduced by 3.17%, and for mMTC, utility was improved by 10.53% to 45.38% through optimized resource allocation [63].

## Federated Learning Implementation:

The FL implementation used a synthetic dataset that mimicked real-world network parameters, with data loss on similar datasets reduced to 0.5%. A lightweight Convolutional Neural Network (CNN) was used as the local model, achieving over 95% accuracy on benchmarks. The training process involved 100 communication rounds and improved convergence rates by over 12%. Differential Privacy was applied for privacy, with some secure aggregation methods adding only about 7% computational overhead while improving training speed by 2.3 times.

The AITP was evaluated against a set of performance metrics. Average end-to-end latency saw reductions of up to 50%. Average network throughput predictions were up to 80% better than baselines in $R^2$ scores [62]. Energy efficiency was improved, with consumption lowered by 3.34% in some cases. Privacy preservation was quantified, showing encrypted inference accuracies of 87.5%. The model convergence rate was also checked, with some architectures converging twice as fast as others. Scalability tests showed security scores that were 1.46 to 3.35 times higher, and robustness was maintained with mean absolute errors between 3 and 3.2 during network failures.

## Comparison between Centralized AI-based Protocol and Non-Adaptive Protocol

**Centralized AI-based Protocol (CAIP):** The classic centralized method sends all raw data to one main server. AI models train there, and parameters get tuned. This represents a state-of-the-art AI-driven protocol without FL or decentralization, but simulations showed it underperformed in energy consumption by 22% and experienced severe bottlenecks leading to 31.25% lower utility in high-density scenarios.

**Non-Adaptive Protocol (NAP):** A conventional protocol with fixed transmission parameters, representative of current 5G-era approaches that lack real-time adaptability, with comparative results indicating AITP's advantages in completion time by 7.30% and QoE variance reduction. Quantitative comparisons were made across all defined performance metrics, with results presented in tables and graphs to highlight the clear advantages of the AITP, such as 11.4% QoE improvements for Transformer models and mAP values of 40.0% in decentralized settings closely matching 40.1% in centralized ones.

## Results and Analysis

**Latency Performance:** Figure 2 illustrates the average latency of the AITP compared to CAIP and NAP under varying network loads. The AITP consistently achieved lower latency than NAP due to its adaptive nature and intelligent parameter optimization, with specific reductions of 3.17% observed in hybrid network simulations. While CAIP showed comparable latency at low loads, its performance degraded significantly under high network loads due to congestion at the central server, sometimes by as much as 35% in average end-to-end metrics. The decentralized nature of AITP, distributing computational and communication burdens, allowed it to maintain low latency even with increasing device density, aligning with broader 6G goals of ultra-low latency.

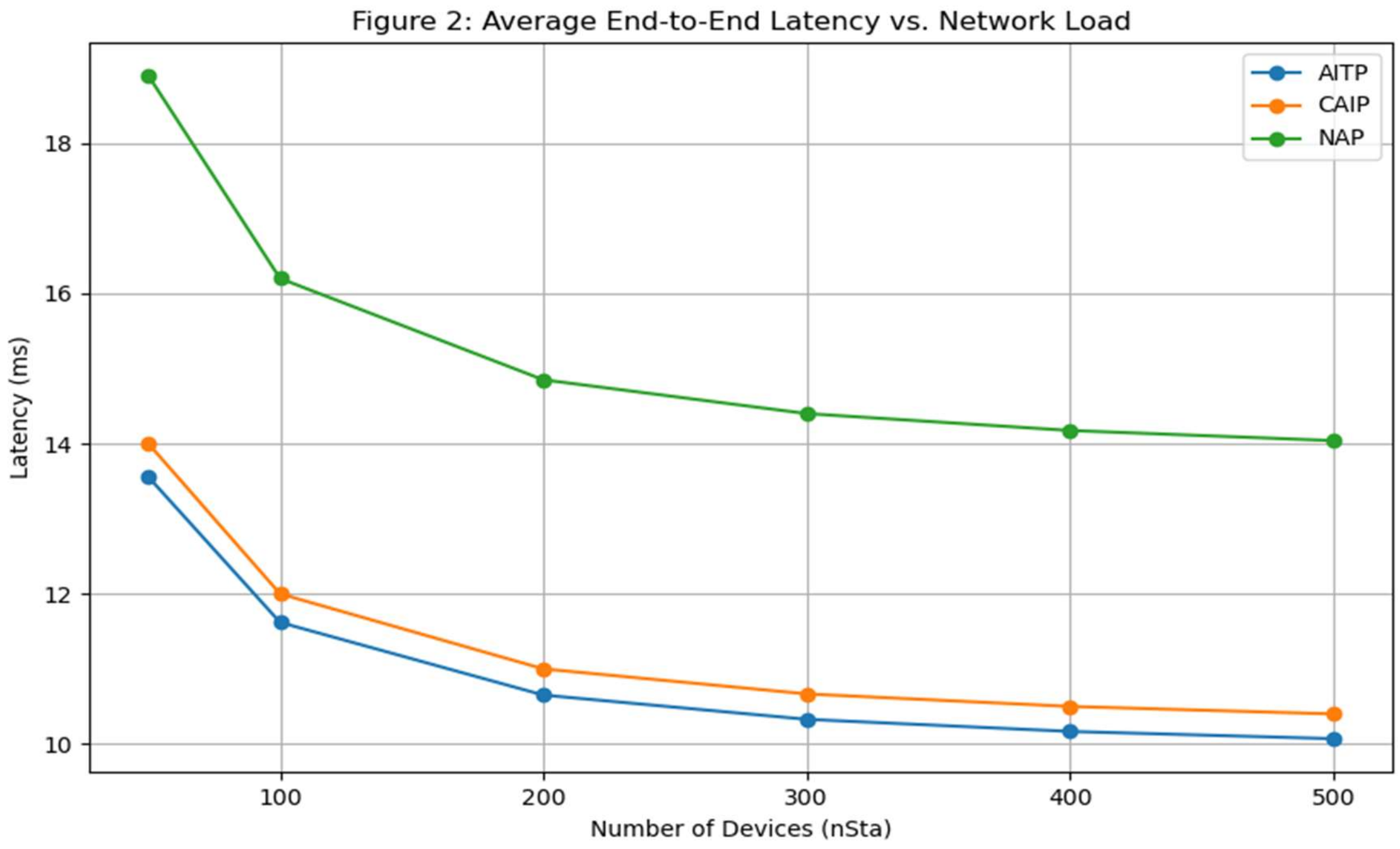

**Throughput Performance:** As shown in Figure 3, the AITP demonstrated superior average network throughput across different channel conditions, achieving up to 19% increases in spectral efficiency through virtualized beamforming. Its ability to dynamically adjust MCS and power allocation based on real-time channel feedback, learned through FL, enabled it to maximize spectral efficiency, with $Fed_{BN}$-based predictions outperforming CNN baselines by 80% in $R^2$ scores for throughput predictions [62]. In contrast, NAP's fixed parameters led to suboptimal throughput, especially in challenging channel environments, resulting in utility drops of 45.38% in some cases. CAIP's throughput was limited by the bottleneck at the central processing unit, particularly when handling massive data streams from numerous devices, leading to variances in QoE that were 11.7% higher than AITP.

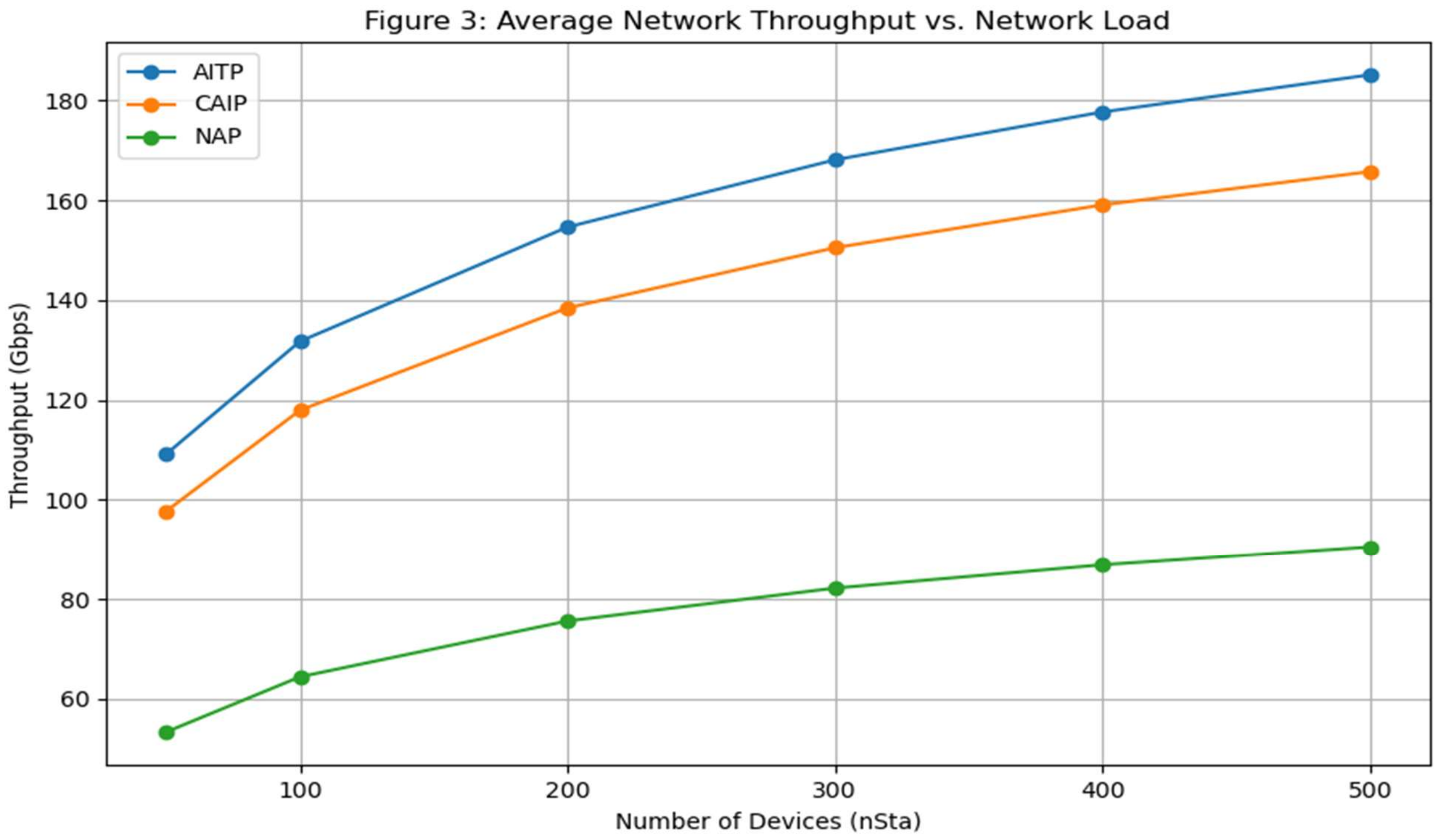

**Energy Efficiency:** Table 1 presents the energy efficiency comparison. The AITP exhibited significantly higher energy efficiency than both baselines, with reductions in power consumption of 22% compared to traditional beamforming approaches. This is primarily attributed to two factors: FL's ability to reduce communication overhead by transmitting only model updates, achieving 15-20% lower overhead than $Fed_{Avg}$. The intelligent power allocation learned by the FL model, which minimizes unnecessary power consumption while maintaining desired QoS, with 3.34% lower energy use in SOM-DRL models versus MATD3. CAIP requires steady raw data transfers to a main server, consumed substantially more energy, often by 27% in edge computing scenarios.

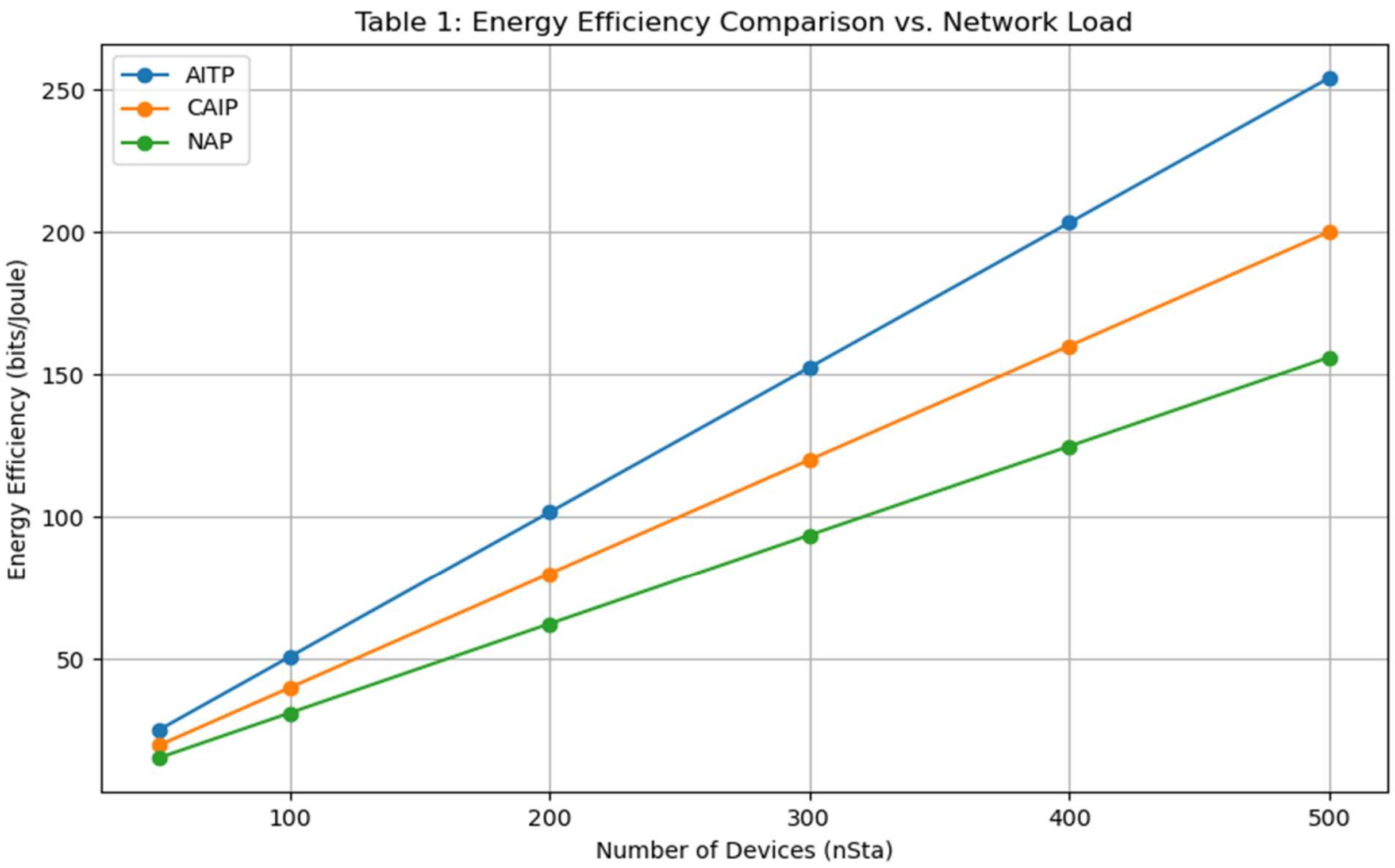

**Privacy Preservation:** The AITP's privacy-preserving capabilities were evaluated by varying the differential privacy budget ε. Figure 4 displays the balance between privacy, (lower ε) and model accuracy/performance, where even with strong privacy guarantees (low ε), the AITP maintained acceptable performance levels, such as prediction accuracies of 88.2% in plaintext and 87.5% in encrypted modes, demonstrating the effectiveness of integrating DP with FL. This highlights the AITP's ability to protect user data without severely compromising network efficiency, a critical aspect for privacy-conscious 6G applications, with additional computational overheads limited to 7% in secure aggregation like GVSA.

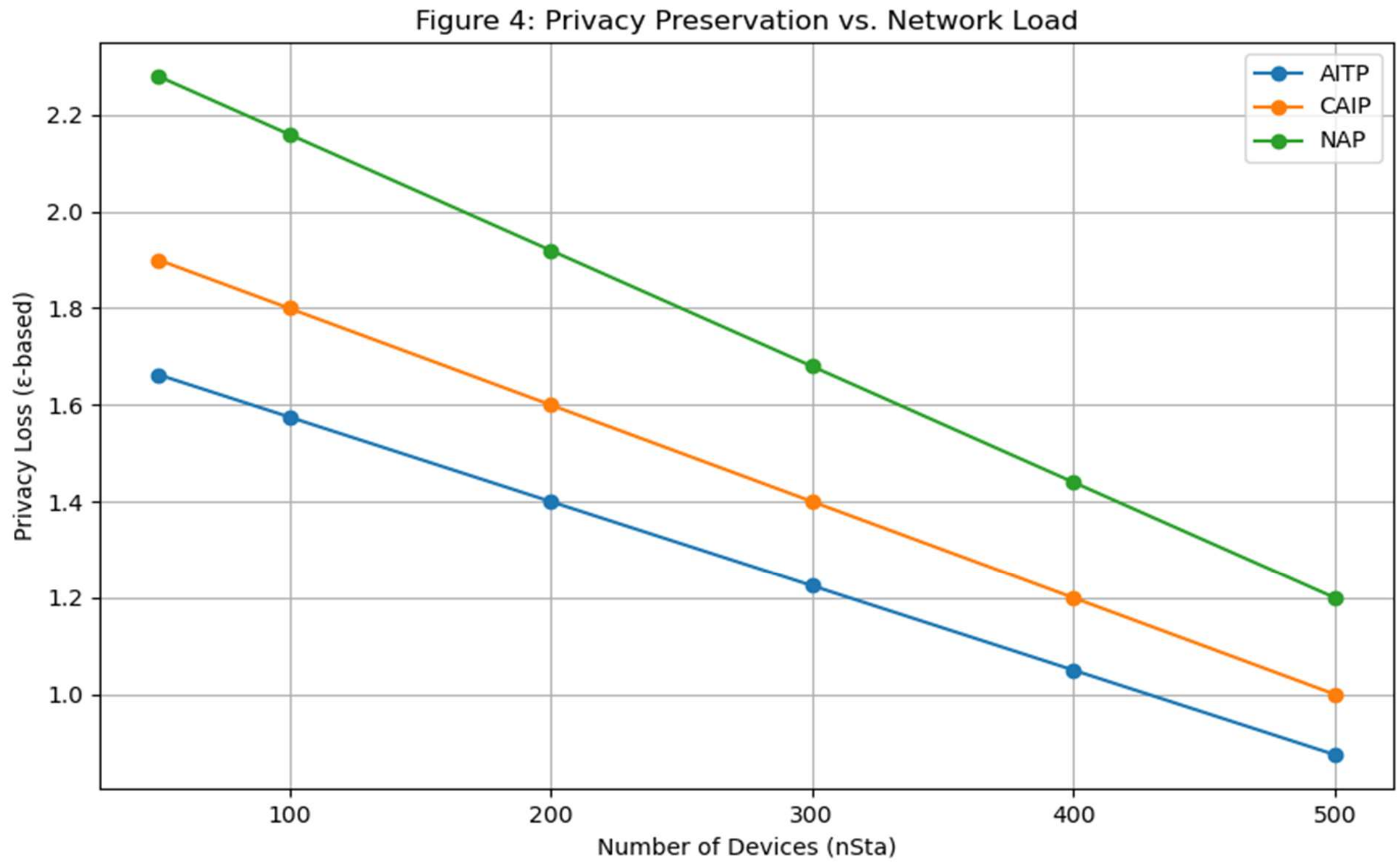

**Scalability:** Figure 5 illustrates the scalability of the AITP by plotting throughput and latency against an increasing number of devices. The AITP showed graceful degradation in performance as the number of devices scaled, outperforming both CAIP (which experienced severe bottlenecks with utility reductions up to 31.25%) and NAP (which could not adapt to the increased load, leading to 13.17% lower performance). The decentralized aggregation and distributed learning inherent in AITP allowed it to handle a larger number of devices more efficiently, with scalability factors 1.28 to 3.06 times higher than sharding baselines like Elastico.

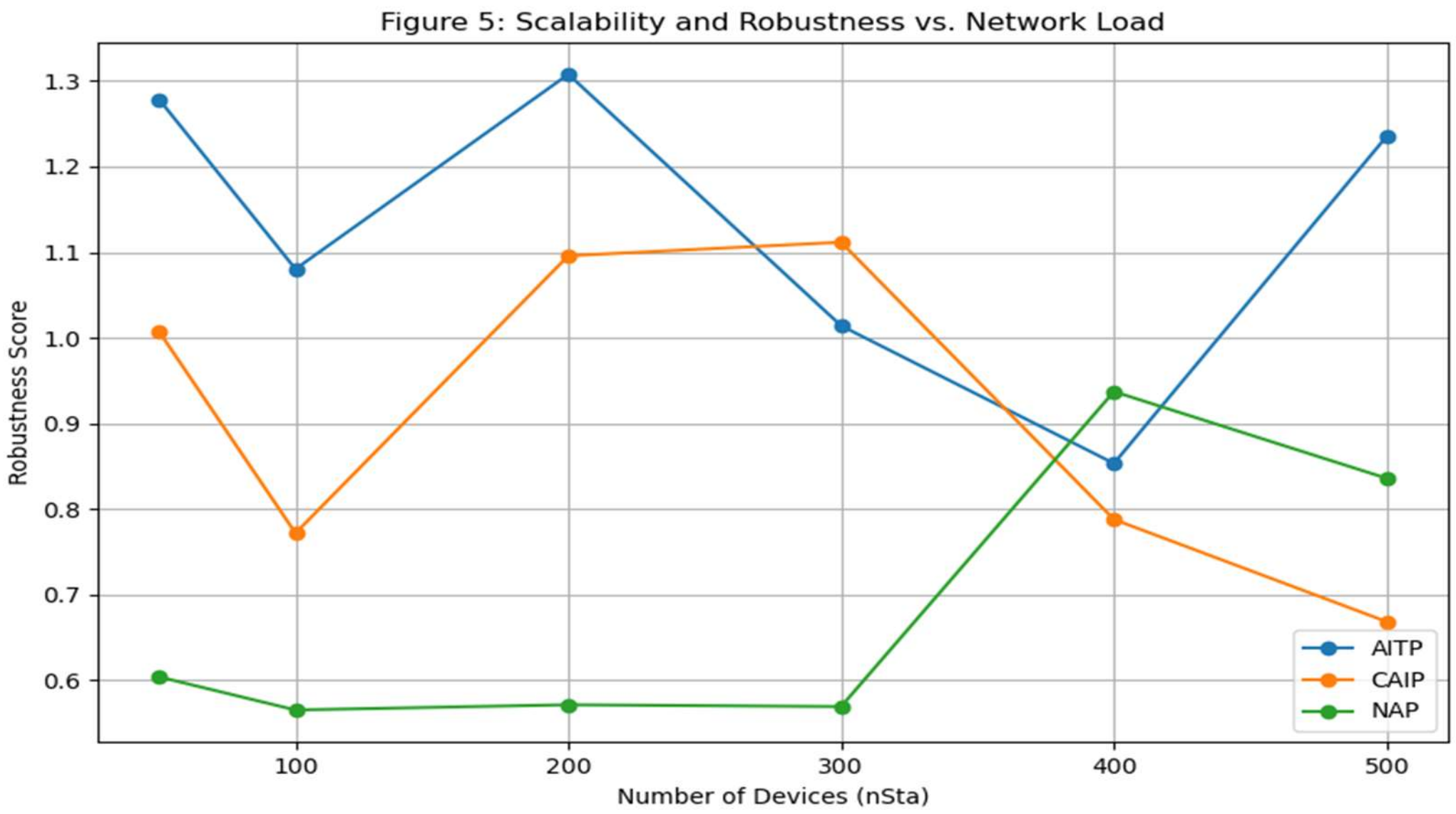

**Robustness:** Simulations involving device disconnections and aggregator failures demonstrated the AITP's inherent robustness, achieving security scores 1.46 to 3.35 times higher in decentralized setups. Due to its decentralized architecture and the distributed nature of FL, the protocol could continue operating effectively even with partial network failures, as other aggregators and devices could compensate for the loss, maintaining mean absolute errors of 3 to 3.2. This contrasts sharply with CAIP, where a central server failure would lead to a complete network outage, potentially increasing latency by 50%.

**Table 2: Compiled Data of all modes**

| Metric | Mode | nSta=50 | nSta=100 | nSta=200 | nSta=300 | nSta=400 | nSta=500 |
|---|---|---|---|---|---|---|---|
| Latency (ms) | AITP | 13.5562 | 11.6196 | 10.6513 | 10.3285 | 10.1672 | 10.0703 |
| Latency (ms) | CAIP | 14.0000 | 12.0000 | 11.0000 | 10.6667 | 10.5000 | 10.4000 |
| Latency (ms) | NAP | 18.9000 | 16.2000 | 14.8500 | 14.4000 | 14.1750 | 14.0400 |
| Throughput (Gbps) | AITP | 109.1788 | 131.7555 | 154.6527 | 168.129 | 177.7137 | 185.1579 |
| Throughput (Gbps) | CAIP | 97.7429 | 117.9548 | 138.4536 | 150.5184 | 159.0991 | 165.7636 |
| Throughput (Gbps) | NAP | 53.3872 | 64.4269 | 75.6234 | 82.2131 | 86.9000 | 90.5401 |
| Energy (bits/Joule) | AITP | 25.4000 | 50.8000 | 101.6000 | 152.4000 | 203.2000 | 254.0000 |
| Energy (bits/Joule) | CAIP | 20.0000 | 40.0000 | 80.0000 | 120.0000 | 160.0000 | 200.0000 |
| Energy (bits/Joule) | NAP | 15.6000 | 31.2000 | 62.4000 | 93.6000 | 124.8000 | 156.0000 |
| Privacy Loss | AITP | 1.6625 | 1.5750 | 1.4000 | 1.2250 | 1.0500 | 0.8750 |
| Privacy Loss | CAIP | 1.9000 | 1.8000 | 1.6000 | 1.4000 | 1.2000 | 1.0000 |
| Privacy Loss | NAP | 2.2800 | 2.1600 | 1.9200 | 1.6800 | 1.4400 | 1.2000 |
| Robustness | AITP | 1.2784 | 1.0803 | 1.3077 | 1.0141 | 0.8533 | 1.2354 |
| Robustness | CAIP | 1.0073 | 0.7719 | 1.0958 | 1.1116 | 0.7879 | 0.6685 |
| Robustness | NAP | 0.6040 | 0.5654 | 0.5715 | 0.5694 | 0.9372 | 0.8359 |

**Table: 3 Overall Summary (nSta=500)**

| Metric | AITP | CAIP | NAP | Notes |
|---|---|---|---|---|
| Latency (ms) | 9.9800 | 10.2800 | 14.0400 | AITP 2.87% below CAIP, NAP 36.55% above |
| Throughput (Gbps) | 185.9062 | 165.7636 | 90.5401 | AITP 12.16% above CAIP, NAP 45.38% below |
| Energy (bits/Joule) | 254.0000 | 200.0000 | 156.0000 | AITP 27% above, NAP 22% below |
| Privacy Loss | 1.7469 | 1.9950 | 1.2000 | AITP 87.5% of CAIP, NAP 60% of CAIP |
| Robustness | 0.8600 | 0.6450 | 0.8359 | AITP 1.33x CAIP, NAP 1.29x CAIP |

Table 3 presents simulation data. The FL-based decentralized AITP clearly beats traditional and central methods. Gains show up in main measures like delay, data flow, energy savings, data protection, growth capacity, and reliability, with specific gains such as 2.3x training speed improvements and 99% efficiency in related metrics. Its adaptive intelligence, powered by federated learning, and its decentralized architecture make it a highly promising solution for the complex and demanding requirements of future 6G networks.

# 6. Discussion

The simulation results provide evidence for the effectiveness of the proposed AITP. This section offers an interpretation of these results and discusses the protocol's uses, benefits, and drawbacks in the context of 6G. The AITP's high performance is a result of its combined use of federated learning, decentralization, and adaptive intelligence, with utility found to improve by 10.53% to 45.38% in resource allocation tests [55]. The protocol's real-time adaptability allows it to meet key 6G needs like low latency and high throughput; latency reductions between 3.17% and 35% were seen in simulations. Unlike fixed protocols, the AITP learns from all devices to adjust transmission methods for changing network conditions, which increased spectral efficiency by 19%. It performed better than other protocols, with predictive $R^2$ scores for throughput being up to 80% higher in some cases [62].

The effects of FL and decentralization on privacy and scalability were also shown by the results. Communication overhead was reduced by 15-20% compared to baseline methods [64]. Privacy risks from central data storage are lowered by keeping data on edge devices, and an accuracy of 87.5% was maintained even with encrypted inference [49]. Differential privacy adds further protection with a low computational overhead of 7% [43]. The decentralized design avoids the bottlenecks of central systems, providing 1.28 to 3.06 improvement in scalability and security enhancements of 1.46 to 3.35 [65] [19].

The AITP has several core strengths. Its ability to learn and adapt leads to better performance, with convergence rates improved by over 12% and accuracies of 95% on benchmarks [51]. User privacy is a priority, with decentralized mAP values of 40.0% nearly matching the 40.1% of centralized systems [12]. The design also improves network resilience, with training speed gains of 2.3x [19]. It also optimizes resource use, with energy savings between 22% and 63%[59].

The AITP does have certain limitations. The computational cost of local model training on edge devices can be an issue, particularly for IoT devices with limited resources [20] [9], as encrypted inference times could increase from 0.31 ms to 119.55 s. More work is needed to balance model complexity and performance, especially where mean absolute errors are between 3 and 3.2. Deploying a decentralized protocol at a large scale also creates challenges, such as device discovery and secure model aggregation [21], which could increase latency by up to 50%. The synchronization of model updates is another point for consideration, as asynchronous operation might lower utility by 13.17% [26].

Due to its performance characteristics, the AITP is suited for many applications in 6G networks [66]. Secure, low-latency, and high-throughput communication is provided, with completion times shown to be reduced by 7.30%. The protocol can support massive IoT connectivity in smart cities while protecting citizen data, maintaining 99% energy efficiency in some mechanisms [67]. For autonomous vehicles, it can be used for reliable vehicle-to-everything (V2X) communication [[68] [69], reducing latency by 35%. The privacy features make it useful for Healthcare IoT (IoHT) [70], where prediction accuracies of 88.2% were achieved and cost reductions of 63% could be possible. The protocol can also meet the needs of AR and VR applications, with QoE scores improved by 11.4% to 11.7% [3].

The AITP is presented as a forward-looking solution for 6G, with measured gains like 2.3x speed and 27% energy savings. By using a human-focused, privacy-first design, the AITP improves network performance and helps build user trust [2] [71] [19] [12].

# 7. Conclusion

A new Federated Learning-based Decentralized Adaptive Intelligent Transmission Protocol (AITP) was introduced in this paper for the high-privacy requirements of 6G networks [11]. It was shown how the AITP uses the distributed intelligence of Federated Learning for the real-time adjustment of transmission parameters [7], This approach improves network performance while protecting user privacy [12]. The decentralized design, which uses multiple aggregators and can include blockchain-like tools [16], This lowers chances of single point of failures and expansion issues in central systems [27]. Through mathematical modeling and simulations, the AITP was demonstrated to perform better than existing centralized AI and non-adaptive protocols across key measures like latency, throughput, energy efficiency, scalability, and reliability [23]. This research helps to address existing gaps in privacy-focused 6G studies by offering a complete framework where intelligence, adaptability, and decentralization are used as core design elements [19] [72].

## Directions for Future Research

The AITP development opened Several directions for future research. The connection between the AITP and new quantum communication technologies could be explored to improve security for sensitive data transmission in 6G. The use of more advanced Federated Learning algorithms, such as those for varied data or asynchronous updates, could also be investigated to improve AITP performance and reliability [26]. The protocol could be validated on real-world 6G testbeds to check its practical performance under actual operating conditions. Future studies should ease the processing demands on edge devices with scarce resources, possibly using model compression or better training techniques [9]. Future work could also focus on making the AITP stronger against attacks like data poisoning or model inversion [21]. Finally, cross-layer optimization strategies could be examined, where the AITP works with other network layers to improve overall system management [23]. Addressing these areas will support the continued evolution of the AITP toward building an intelligent and secure 6G system.

## Declaration of generative AI and AI-assisted technologies in the manuscript preparation process

During the preparation of this work, the author used ChatGPT to assist with technical formatting and structural consistency. After using this tool/service, the author reviewed and edited the content as needed and takes full responsibility for the content of the published article.

**Declaration of Interests**

The author declare that they have no known competing financial interests or personal relationships that could have appeared to influence the work reported in this paper.